\RequirePackage{lineno}
 
\documentclass[5p,twocolumn]{elsarticle}
\usepackage{graphicx}

\begin{document}

\title{\rightline{\small IIT-CAPP-10-4}\vspace{-.125in}
\rightline{\small FERMILAB-PUB-10-260}Study of the Rare Hyperon Decay
${\boldmath
\Omega^\mp \to \Xi^\mp \: \pi^+\pi^-}$
}

\date{\today}

\author[iit]{{\large The HyperCP Collaboration:}\\[.1in]
O. Kamaev\fnref{fn1}
}
\author[iit]{N.\,Solomey\fnref{fn2}}
\author[iit]{R.\,A.\,Burnstein}
\author[iit]{A.\,Chakravorty\fnref{ac-addr}}
\author[as]{Y.\,C.\,Chen}
\author[ucb]{W.\,S.\,Choong\fnref{wc-addr}}
\author[usa]{K.\,Clark}
\author[uva]{E.\,C.\,Dukes}
\author[uva]{C.\,Durandet\fnref{cd-addr}}
\author[ug]{J.\,Felix}
\author[lbnl]{Y.\,Fu}
\author[lbnl]{G.\,Gidal}
\author[umich]{H.\,R.\,Gustafson}
\author[uva]{T.\,Holmstrom\fnref{th-addr}}
\author[uva]{M.\,Huang\fnref{mh-addr}}
\author[fnal]{C.\,James}
\author[usa]{C.\,M.\,Jenkins}
\author[lbnl]{T.\,D.\,Jones}
\author[iit]{D.\,M.\,Kaplan\corref{ca}}
\ead{kaplan@iit.edu}
\author[umich]{M.\,J.\,Longo}
\author[uva]{L.\,C.\,Lu\fnref{ll-addr}}
\author[iit]{W.\,Luebke}
\author[ucb,lbnl]{K.\,B.\,Luk}
\author[uva]{K.\,S.\,Nelson\fnref{kn-addr}}
\author[umich]{H.\,K.\,Park\fnref{hp-addr}}
\author[ul]{J.-P.\,Perroud}
\author[iit]{D.\,Rajaram\fnref{dr-addr}}
\author[iit]{H.\,A.\,Rubin}
\author[fnal]{J.\,Volk}
\author[iit]{C.\,G.\,White}
\author[iit]{S.\,L.\,White\fnref{sw-addr}}
\author[lbnl]{P.\,Zyla}
\address[as]{Institute of Physics, Academica Sinica, Taipei 11529, Taiwan, Republic of China}
\address[ucb]{University of California, Berkeley, California 94720, USA} 
\address[lbnl]{Lawrence Berkeley National Laboratory, Berkeley, California 94720, USA}
\address[fnal]{Fermi National Accelerator Laboratory, Batavia, Illinois 60510, USA}
\address[ug]{University of Guanajuato, 3700 Leon, Mexico}
\address[iit]{Illinois Institute of Technology, Chicago, Illinois 60616, USA}
\address[ul]{University of Lausanne, CH-1015 Lausanne, Switzerland}
\address[umich]{University of Michigan, Ann Arbor, Michigan 48109, USA}
\address[usa]{University of South Alabama, Mobile, Alabama, 36688, USA}
\address[uva]{University of Virginia, Charlottesville, Virginia 22904, USA}
\fntext[fn1]{Present address: Univ.\ of Minnesota, Minneapolis, MN 55455, USA}
\fntext[fn2]{Present address: Wichita State Univ., Wichita, KS 67260, USA}
\fntext[ac-addr]{Present address: St.\ Xavier Univ., Chicago, IL 60655, USA}
\fntext[wc-addr]{Present address: Lawrence Berkeley National Laboratory, Berkeley, CA 94720, USA}
\fntext[cd-addr]{Present address: Paradise Valley College, Phoenix, AZ  85032, USA}
\fntext[th-addr]{Present address: Longwood Univ., Farmville, VA 23909, USA}
\fntext[mh-addr]{Present address: Leung Center for Cosmology and Particle Astrophysics, National Taiwan Univ., Taipei, 10617 Taiwan}
\fntext[ll-addr]{Present address: Coll.\ of Medicine, Ohio State Univ., Columbus, OH 43210, USA}
\fntext[kn-addr]{Present address: Applied Physics Laboratory, Johns Hopkins Univ., Laurel, MD 20723, USA}
\fntext[hp-addr]{Present address: Kyungpook National Univ., Daegu 702-701, Korea}
\fntext[dr-addr]{Present address: Univ.\ of Michigan, Ann Arbor, MI 58109, USA}
\fntext[sw-addr]{Present address: Radiology Dept., Univ.\ of Alabama, Birmingham, AL 35249, USA}
\cortext[ca]{Corresponding author}

\begin{abstract}
We report
a new measurement of the decay $\Omega^- \to \Xi^-\pi^+\pi^-$ with 76 events and a first observation of the decay $\overline{\Omega}^+\to  \overline{\Xi}^+ \pi^+ \pi^-$ with 24 events, yielding a combined branching ratio $(3.74 ^{+0.67}_{-0.56}) \times 10^{-4}$. This represents a factor 25 increase in statistics over the best previous measurement.
No evidence is seen for {\em CP} violation, with ${\cal B}(\Omega^- \to \Xi^-\pi^+\pi^-)=4.04^{+0.83}_{-0.71} \times10^{-4}$ and ${\cal B}({\overline \Omega}{}^+ \to {\overline \Xi}{}^+\pi^+\pi^-)=3.15^{+1.12}_{-0.89} \times10^{-4}$.
Contrary to theoretical expectation, we see little evidence for the decays $\Omega^-\to\Xi_{1530}^{*0}\pi^-$ and ${\overline\Omega}{}^+\to{\overline \Xi}_{1530}^{*0}\pi^+$ and place a 90\% C.L. upper limit on the combined branching ratio ${\cal B}(\Omega^-({\overline\Omega}{}^+) \to \Xi^{*0}_{1530}({\overline  \Xi}{}^{*0}_{1530}) \pi^\mp)<7.0\times10^{-5}$. 
\end{abstract}

\maketitle

	Although the $\Omega^-$ hyperon was discovered in 1964, many of its decay modes are still poorly known experimentally. One of these is $\Omega^-  \rightarrow  \Xi^- \pi^+ \pi^-$. The $\Omega^-$ is the only ground-state hyperon massive enough to have such a $\Delta S = 1$, three-body, nonleptonic weak decay. This decay has long been assumed by theorists to proceed predominantly via the $\Xi_{1530}^{*0}$ intermediate state~\cite{gs,fg}. Based on four observed $\Omega^-  \rightarrow  \Xi^- \pi^+ \pi^-$ events, Bourquin {\it et al.}\ established the best previous measurement of the branching ratio: ${\cal B}(\Omega^-  \rightarrow  \Xi^- \pi^+ \pi^-)= 4.3^{+3.4}_{-1.3}\times10^{-4}$~\cite{wa2}. However, they were unable to distinguish the nonresonant channel from the resonant $\Omega^-\to\Xi^{*0}_{1530} \pi^- \to \Xi^-\pi^+\pi^-$ decay. In interpreting this observation, the Particle Data Group (PDG) adopted the assumption by Finjord and Gaillard~\cite{fg} that  the  $\Xi^-\pi^+\pi^-$ final state in $\Omega^-$ decay arises entirely from the $\Xi^{*0}_{1530}$, hence
that ${\cal B}(\Omega^-  \rightarrow  \Xi^- \pi^+ \pi^-)= \frac{2}{3}\,{\cal B}(\Omega^-  \rightarrow \Xi^{*0}_{1530} \pi^-)$ by isospin symmetry, giving the estimate ${\cal B}(\Omega^-  \rightarrow \Xi^{*0}_{1530} \pi^-)= 6.4^{+5.1}_{-2.0}\times10^{-4}$~\cite{pdg}. Yet, until now, that assumption has not been tested by experiment.

\begin{figure}
\includegraphics[width=\columnwidth]{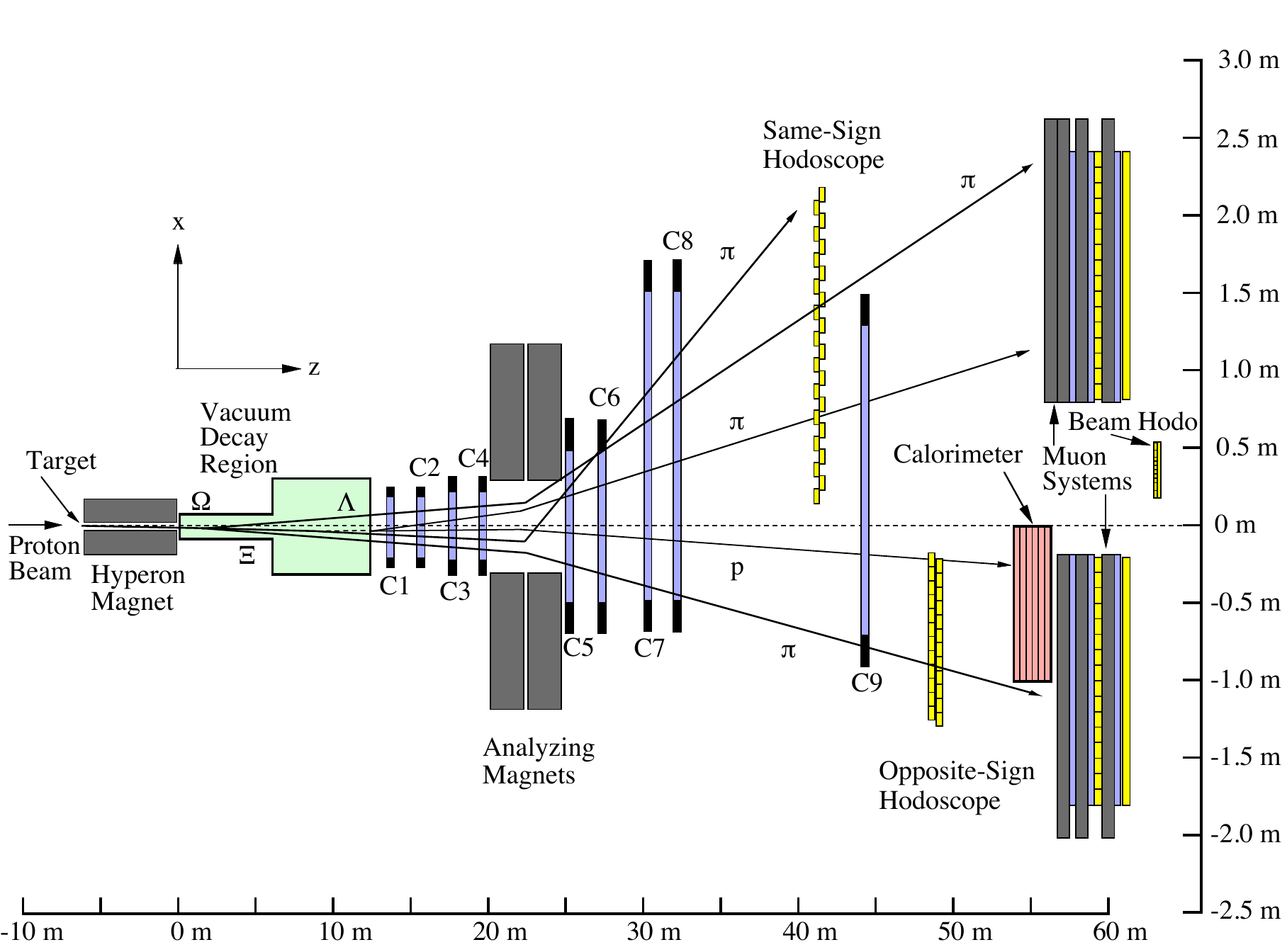}
\caption{Plan view of the HyperCP apparatus. The typical topology of the signal decay
$\Omega^- \to\Xi^- \pi^+\pi^- \to(\Lambda\pi^-) \pi^+ \pi^- \to (p \pi^- \pi^-) \pi^+ \pi^-$  is shown.  }
\label{e871}
\end{figure}

The HyperCP experiment (Fermilab E871) took data in 1997 and 1999.
The apparatus, depicted schematically in Fig.~\ref{e871}, is described in detail elsewhere \cite{nim}. In brief, an 800\,GeV/$c$ proton beam struck a target located at the entrance of a collimating channel 
within the ``hyperon" dipole magnet. The selected momentum  ranged from 110 to 240\,GeV/$c$, with an average of $\approx$\,160 GeV/$c$ and r.m.s.\ spread of $\approx$\,25 GeV/$c$. A 13-meter-long evacuated decay volume started at the exit of this hyperon channel and was viewed by a charged-particle tracking spectrometer 
comprising four multiwire proportional chambers preceding and five following a pair of dipole analyzing magnets. 
The polarities of the magnets were such that particles having  the same charge as the beam or opposite to it were deflected to the left or right, respectively. 
The tracking chambers were followed by trigger hodoscopes in each arm of the spectrometer. Also used in the trigger was a hadron calorimeter in the ``opposite-sign" arm. For the data sample discussed here, the trigger requirement was at least 60 GeV deposited energy in the calorimeter and hits in the hodoscopes consistent with the passage of at least two charged particles of opposite polarities.  
Antihyperon processes were studied using data recorded under identical trigger conditions (referred to below as ``positive'' data) to those for hyperons by reversing the magnetic-field directions of both the hyperon and analyzing magnets;  singles rates  in the spectrometer were approximately equalized by using a target one-third as long for positive data.

The experiment recorded $\sim$30,000 5\,GB 
data tapes containing some $2.3\times10^{11}$ events. These raw tapes were processed 
\cite{chep} 
to produce output data sets satisfying various topological or particle-ID requirements depending upon their intended use.
The selection cuts for the analysis presented here were initially devised from Monte Carlo (MC)-simulated data
samples for signal ($\Omega^- \rightarrow \Xi^- \pi^+ \pi^-$, $\Xi^{-}\to\Lambda\pi^{-}$, $\Lambda\to p\pi^-$) and normalizing ($\Omega^- \rightarrow \Lambda K^-$, $\Lambda\to p\pi^-$, $K^- \to \pi^-\pi^+\pi^-$)
modes, prior to examination of the data samples.
The cuts used in this analysis required: 
(a) at least three negative and two
positive tracks (or, for the positive data sample,
three positive and two negative tracks)~\cite{antiparticles},
(b) $\Omega$, $\Xi$, $\Lambda$, and $K$ decay vertices located inside the decay
volume, (c) a topology consistent with the intended decay, 
(d) a combined 5-track momentum between 135 and 220 GeV/$c$, 
(e) reconstructed $\Omega$, $\Xi$, $\Lambda$, and $K$
masses within $\pm 3$ standard deviations ($\sigma$) of their nominal values~\cite{pdg}, and (f) that the
reconstructed $\Omega$ track extrapolate to the target and pass through the aperture of the collimator. By Monte Carlo simulation we find the overall efficiencies of these cuts to be 50\% for the signal mode and 43\% for the normalizing mode~\cite{thesis}.
The MC simulation was carefully checked against data and successfully reproduced mass resolutions and particle spatial and momentum distributions. 

In each event, the opposite-sign track with the highest momentum was assumed to be the proton. MC simulation showed this assumption to be correct for more than 99\% of events. This assumption also gave a reconstructed invariant $\Omega$ mass closest to the PDG value, when trying all possible track combinations, for 100\% of data events with five charged tracks and for 98\% of events with more than five charged tracks (which constituted $\approx$\,35\% of the data sample). In determining which same-sign track to associate with the proton to form a $\Lambda$, that pion resulting in the best $\Lambda$ mass (i.e., closest to the PDG 2008  average value of 1,115.68\,MeV/$c^2$~\cite{pdg}) was chosen, and similarly for the $\Xi$. 
For normalizing-mode events with more than 5 tracks, after reconstruction of the $\Lambda$, the three remaining tracks of the appropriate polarities giving the best $K$ mass were used.

We normalized the candidate events using the $\Omega^- \rightarrow \Lambda K^-$, $\Lambda\to p \pi^-$, $K^- \to \pi^-\pi^+\pi^-$ decay since its topology is similar to that of the signal mode. 
The MC was used to calculate the HyperCP apparatus acceptance corrections for the normalizing and signal modes. 
The signal mode was simulated with a uniformly populated phase-space generator.

\begin{figure*}
\vspace{0.1in}
\includegraphics[width=.495\linewidth,trim=10 15 25 25 mm]{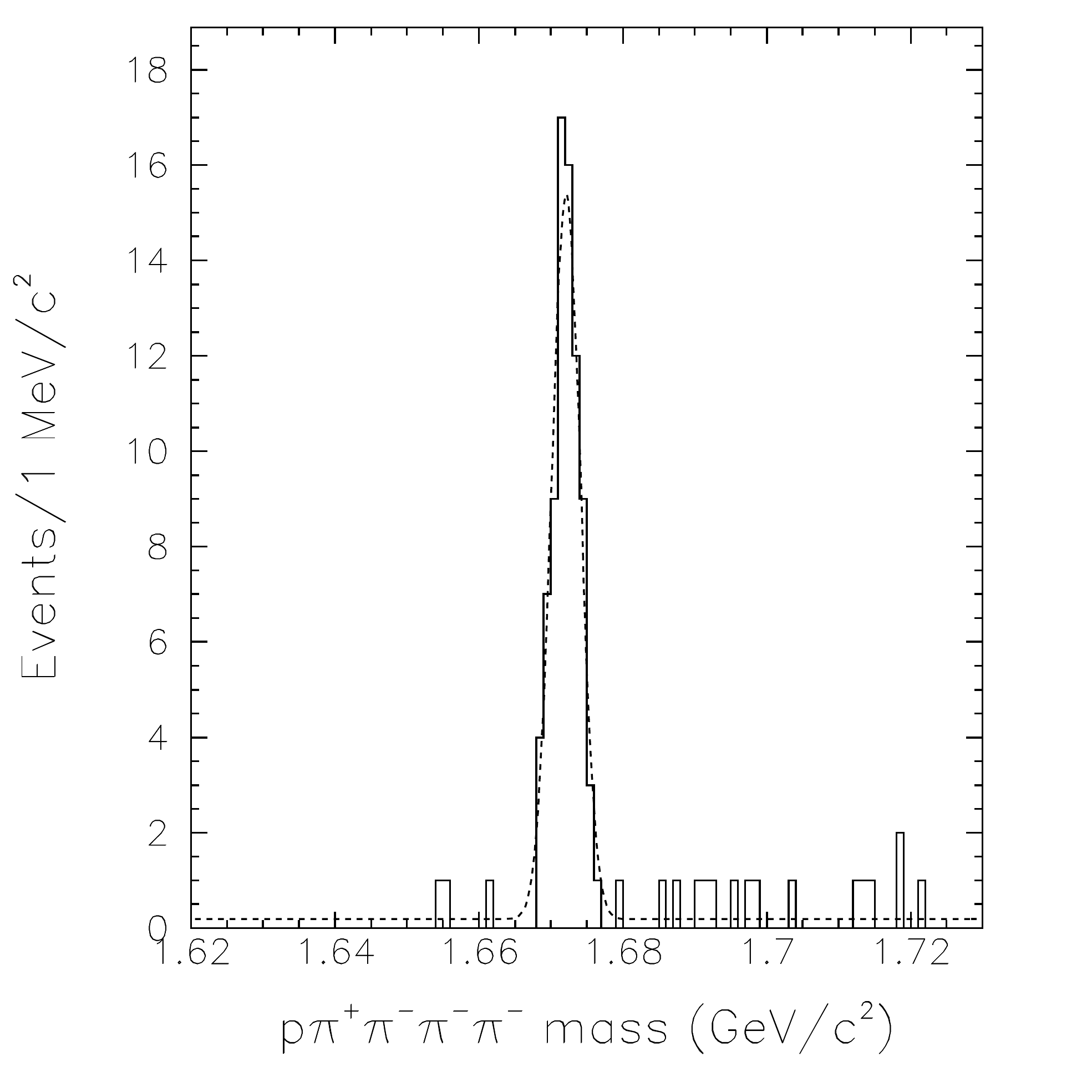}
\includegraphics[width=.495\linewidth,trim=10 15 25 25 mm]{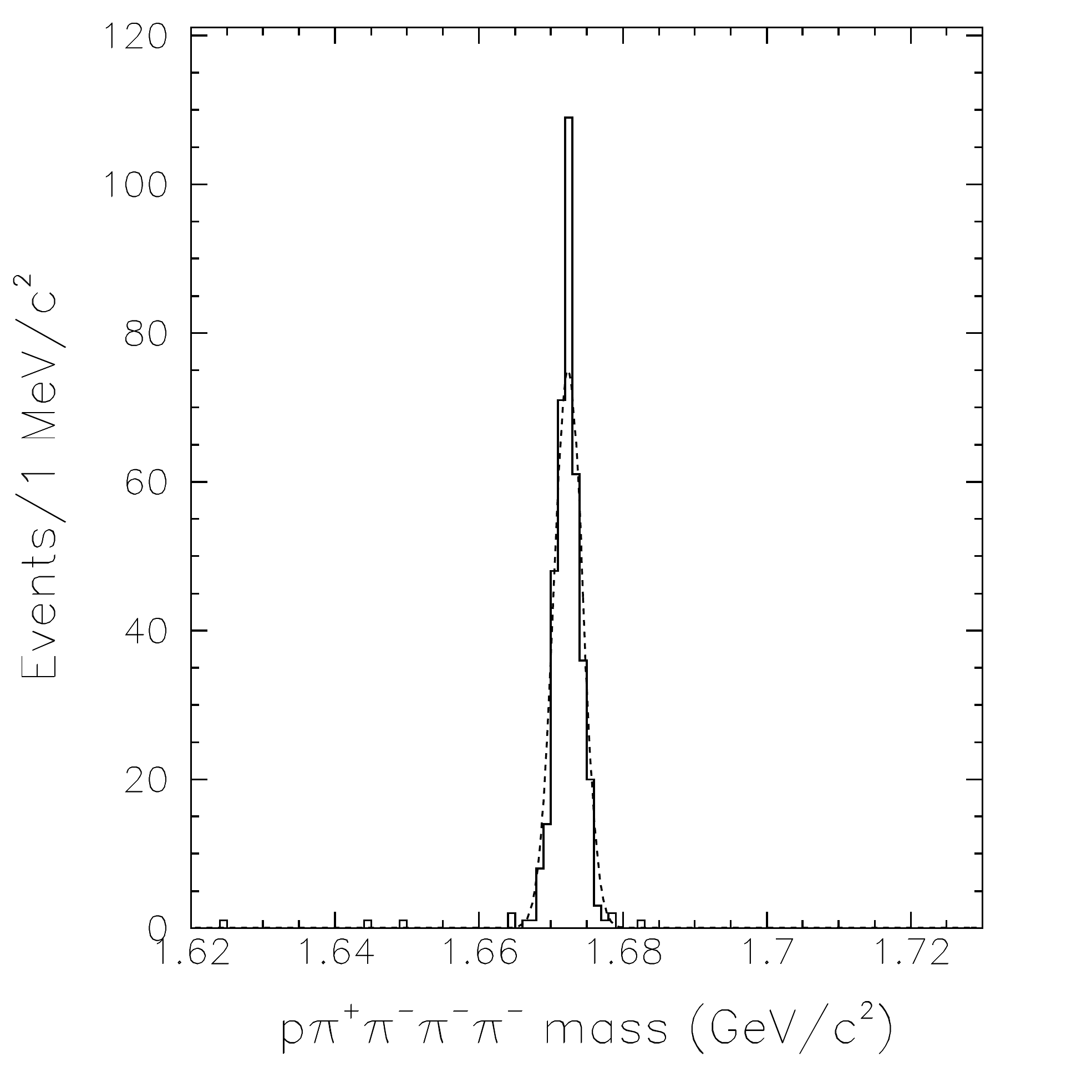}
\caption{Observed-event histograms for (left)  $\Omega^- \to \Xi^-\pi^+\pi^-$ signal mode and (right) normalizing mode $\Omega^-\rightarrow \Lambda K^-$ where the $\Lambda$ decays into $p \pi^-$ and the $K^-$ decays into three charged pions; dashed curves show Gaussian-plus-constant fits to the data as described in text.}
\label{data-minus}
\end{figure*}

\begin{figure*}
\includegraphics[width=.495\linewidth,trim=10 15 25 25 mm]{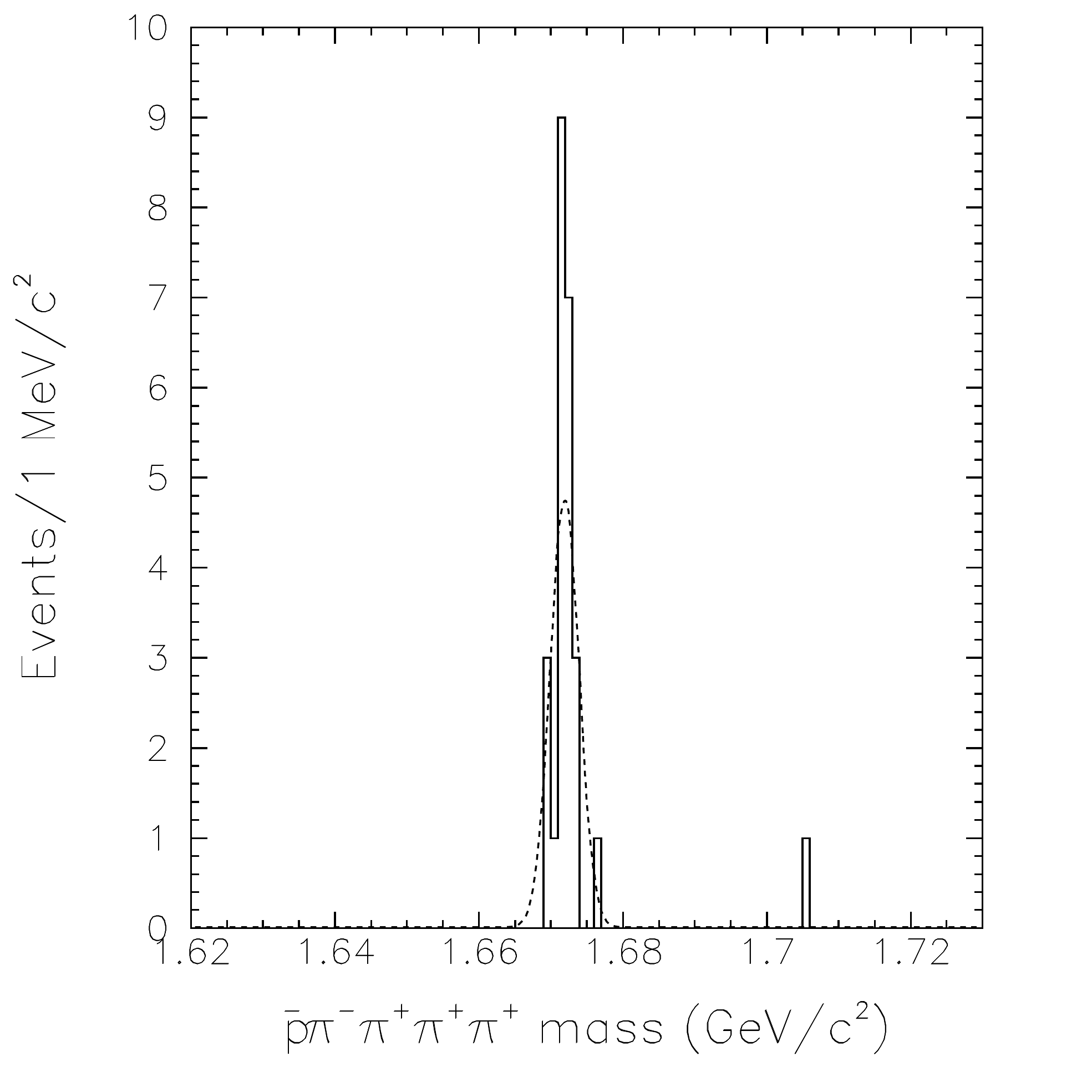}
\includegraphics[width=.495\linewidth,trim=10 15 25 25 mm]{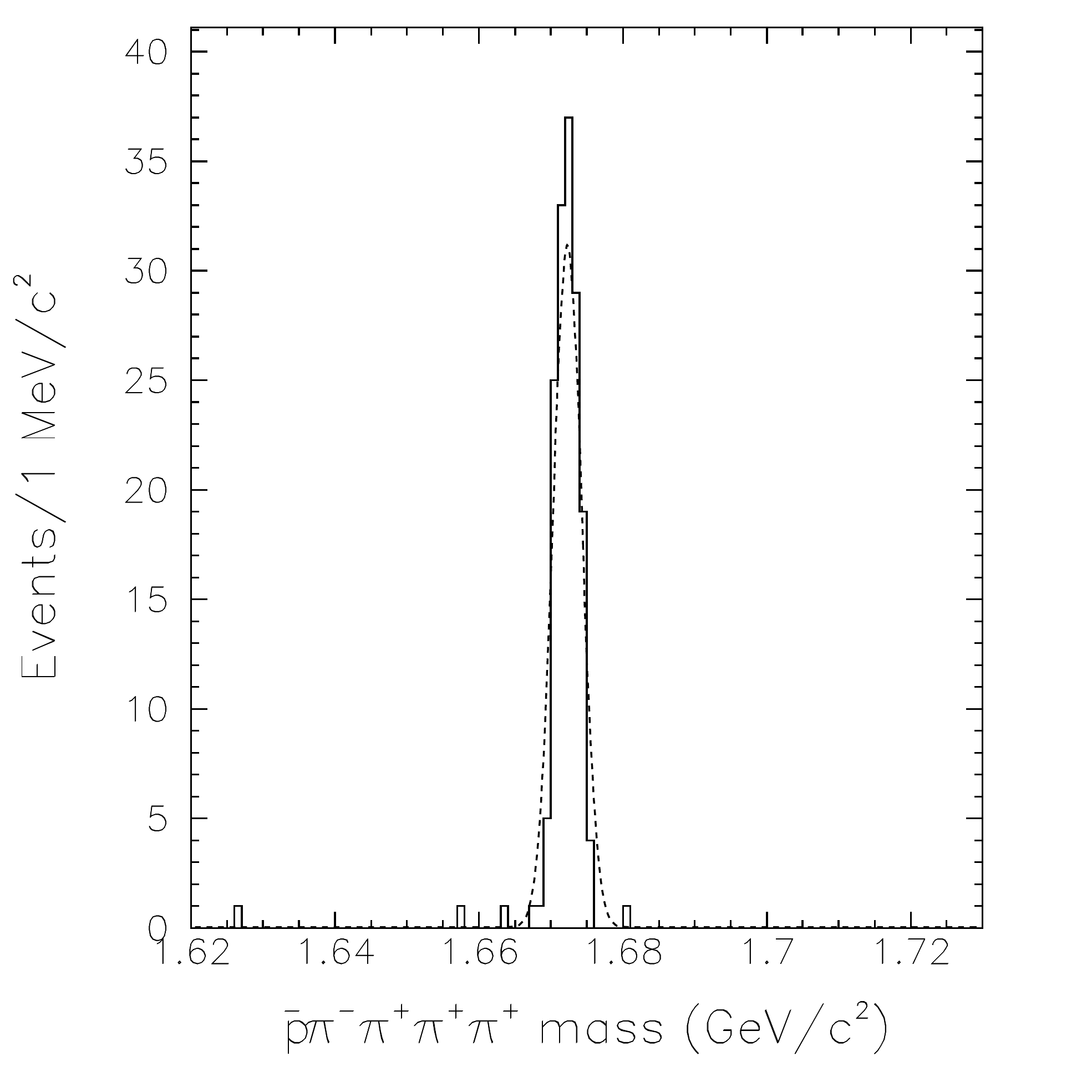}
\caption{Observed-event histograms for (left) ${\overline \Omega}{}^+ \to {\overline \Xi}{}^+\pi^+\pi^-$ 
signal mode and (right) normalizing mode ${\overline \Omega}{}^+\rightarrow {\overline \Lambda} K^+$ where the $\overline \Lambda$ decays into ${\overline p} \pi^+$ and the $K^+$ decays into three charged pions; dashed curves show Gaussian-plus-constant fits to the data as described in text.}
\label{data-plus}
\end{figure*}

Results based on the larger, 1999 HyperCP data sample are reported here. The observed
negative (positive) five-track mass distributions of the signal and normalizing decays 
are shown in Fig.~\ref{data-minus} (Fig.~\ref{data-plus}); we observe 78 (24) events within $\pm3$$\sigma$ of the MC mass resolution ($\sigma=2.0$\,MeV/$c^2$).
The unbinned log-likelihood fits shown in the figures employ a Gaussian for the mass peak, with r.m.s.\ width (2.0\,MeV/$c^2$) determined by fitting the MC signal and normalization samples, plus a constant representing the background.  From the fits, we estimate $2.3\pm0.5$ ($0.1\pm0.1$) background events under the peak, within $\pm3\sigma$, for the negative (positive) signal mode.
For the normalizing mode,  375 (156) events are observed within $\pm$3$\sigma$ of the MC mass resolution including an estimated background of $0.5\pm 0.3$ ($0.4\pm 0.2$) events.


\begin{figure*}
\vspace{0.1in}
\includegraphics[width=.4975\linewidth]{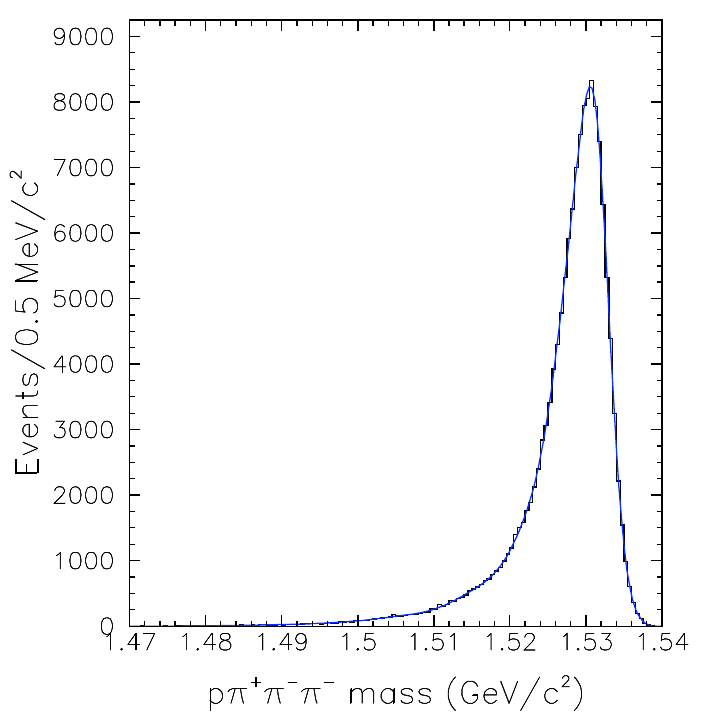}
\includegraphics[width=.5\linewidth]{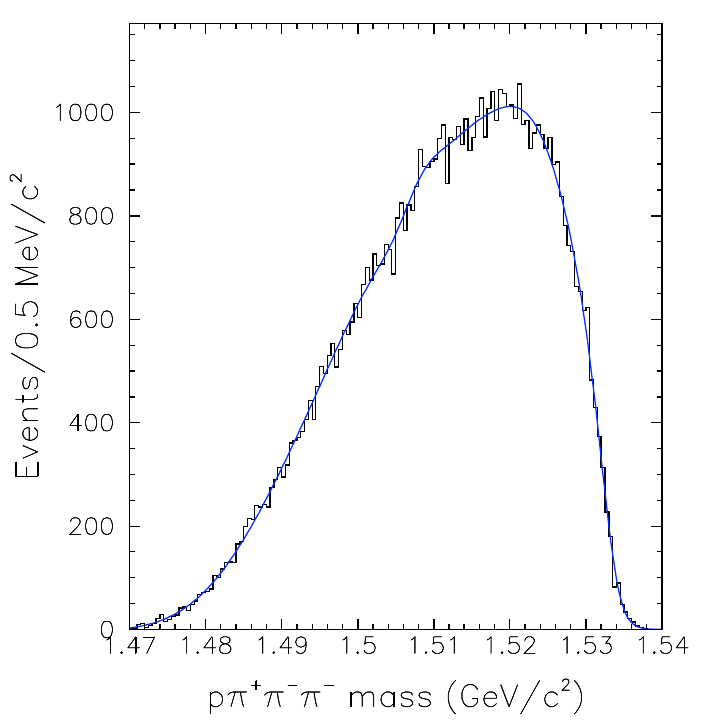}
\caption{Mass distribution of $p \pi^+ \pi^- \pi^-$  for (left) resonance-mode and (right) direct  decays from MC simulation; curves are analytical functions determined by smoothing these histograms.}
\label{fit-func}
\end{figure*}

\begin{figure*}
\includegraphics[width=.5\linewidth]{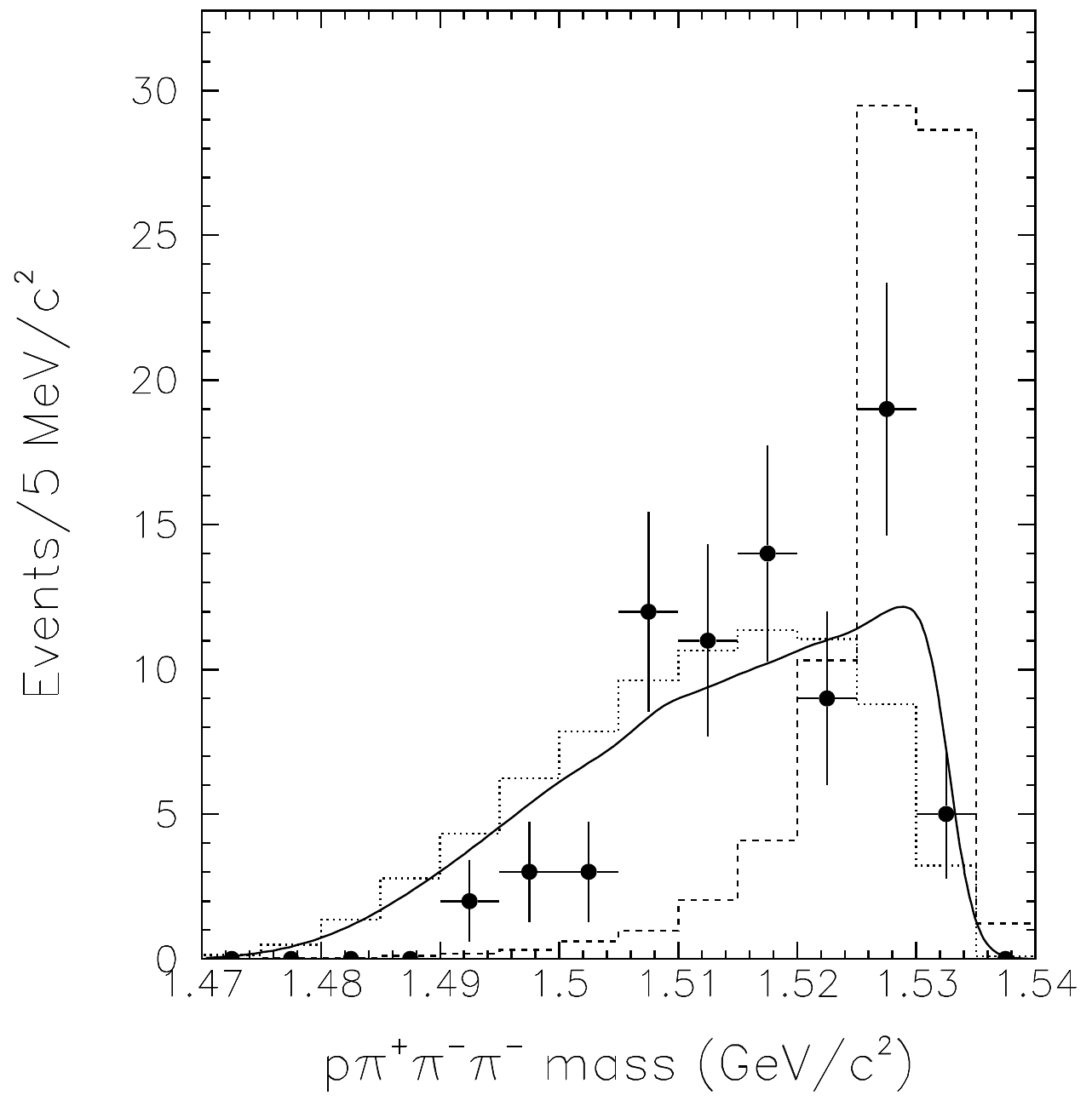}
\includegraphics[width=.5\linewidth]{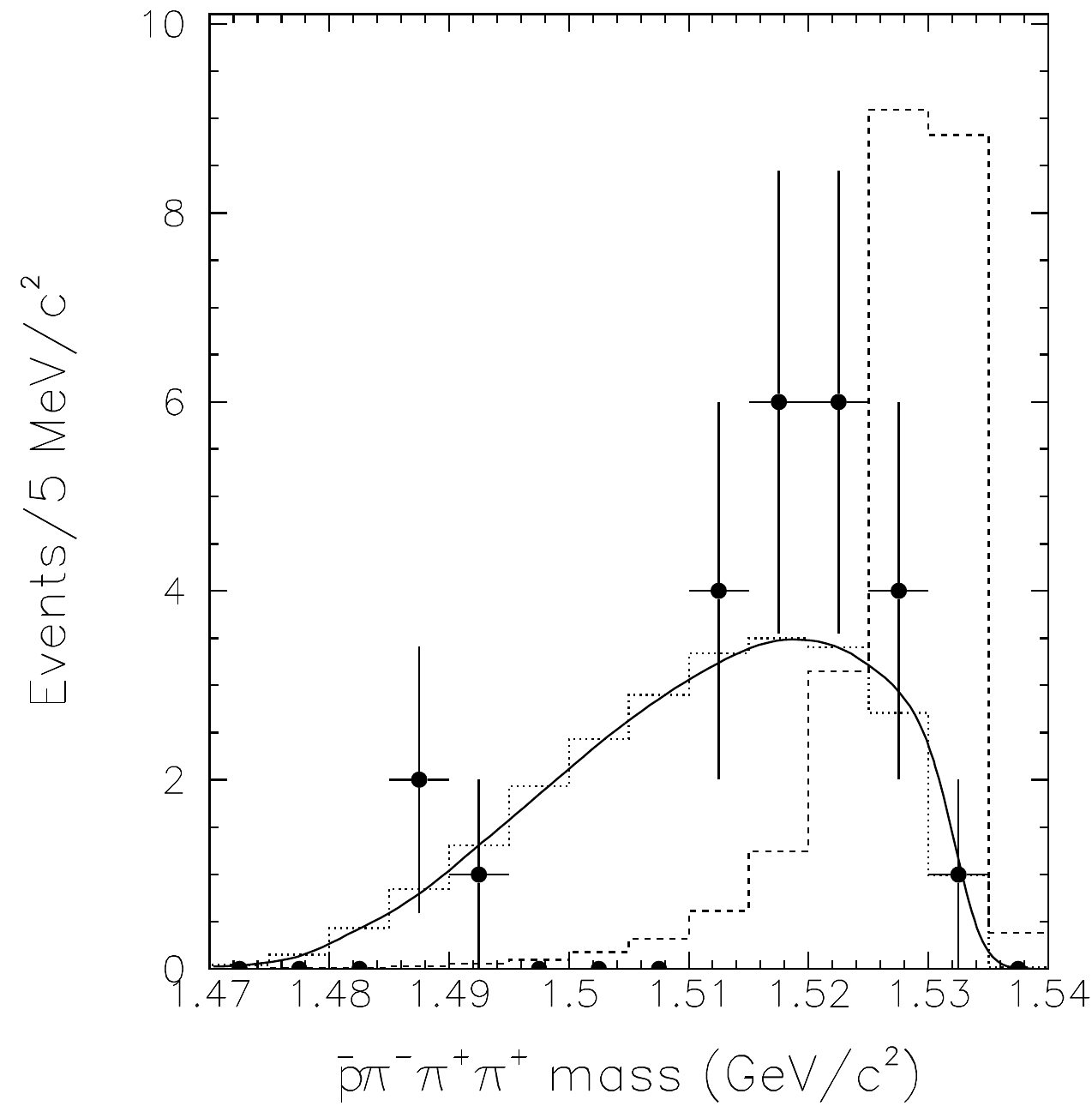}
\caption{Mass distributions of (left) $p \pi^+ \pi^- \pi^-$ and  (right) $\overline{p} \pi^- \pi^+ \pi^+$, together with fits (solid curves) to sums of resonance-mode (dashed histograms) and direct (dotted histograms) functional forms; dashed and dotted histograms are each normalized to the total number of events.}
\label{mass-fits}
\end{figure*}

The signal events presumably represent a combination of direct  $\Omega^- \rightarrow \Xi^- \pi^+\pi^-$ decays with branching ratio ${\cal B}_{dir}$ and those proceeding via the resonant decay mode  $\Omega^- \rightarrow\Xi_{1530}^{*0}\pi^-$  with branching ratio ${\cal B}_{res}$, with the subsequent decay $\Xi_{1530}^{*0}\rightarrow\Xi^- \pi^+$. 
Because of the short lifetime of the $\Xi^{*0}_{1530}$, the $\Xi^{*0}_{1530}$ and $\Xi^-$ vertices are indistinguishable. The main difference between the resonant and direct modes that we have identified in MC studies is the respective distributions in $\Xi^-\pi^+$ invariant mass: as shown in Fig.~\ref{fit-func}, the resonance-mode mass distribution is sharply peaked at the resonance mass, while the direct one is more broadly peaked at $\approx 1.52$\,GeV/$c^2$. 
We  decompose the $N_{sig}$ signal events into numbers of direct ($N_{dir}$) and resonance ($N_{res}$) events according to
\begin{equation}
N_{dir}\equiv N_{\Omega^- \rightarrow \Xi^- \pi^+\pi^-} =   f_{dir} \times N_{sig}\,,\\
\end{equation} 
\begin{equation} 
N_{res} \equiv N_{\Omega^- \rightarrow \Xi_{1530}^{*0}\pi^-\to \Xi^- \pi^+\pi^-} =  f_{res} \times N_{sig}\,. 
\end{equation} 
To determine the resonance ($f_{res}$) and direct ($f_{dir}$) fractions we fit the signal-event $\Xi^-\pi^+$ mass distributions with a linear combination of the functional forms shown in Fig.~\ref{fit-func}. Results are summarized in Table~\ref{fit-results};  the fit functions together with corresponding experimental $\Xi^{-}\pi^{+}$ invariant mass distributions are plotted in Fig.~\ref{mass-fits}. Due to the small numbers of observed events, the fits were performed using the unbinned generalized log-likelihood technique~\cite{Frodesen}. The goodness of fit was evaluated using the $\chi^2$ and the Kolmogorov test and was acceptable in all cases~\cite{thesis}. No significant resonance component was observed. 

We calculate 
\begin{eqnarray}\    
{\cal B}_{res}\equiv{\cal B}(\Omega^{-} \to \Xi^{*0}_{1530}
    \pi^{-}) = f_{res}
\frac{N_{sig}} {N_{norm}} \frac{A_{norm}}{A_{res}}\nonumber \\\label{eq:Bres}
    \times
    \frac{{\cal B}_{\Omega^{-} \to
    \Lambda K^{-}}\cdot {\cal B}_{K^{-}\to \pi^{+}\pi^{-}\pi^{-}}}
    {{\cal B}_{\Xi^{*0}_{1530} \to \Xi^{-} \pi^{+}}\cdot
     {\cal B}_{\Xi^{-}\to\Lambda\pi^{-}}},
\end{eqnarray}
and also obtain signal-mode branching fractions for the direct process   
\begin{eqnarray}
{\cal B}_{dir}(\Omega^{-} \to \Xi^{-} \pi^{+}\pi^{-}) = f_{dir}
\frac{N_{sig}} {N_{norm}} \frac{A_{norm}}{A_{dir}}\nonumber \\
    \times
    \frac{{\cal B}_{\Omega^{-} \to
    \Lambda K^{-}}\cdot {\cal B}_{K^{-}\to \pi^{+}\pi^{-}\pi^{-}}}
    {{\cal B}_{\Xi^{-}\to\Lambda\pi^{-}}},\label{eq:BR}
\end{eqnarray}
where $N$ is the number of events and $A$ is the acceptance for the indicated mode as derived from MC simulation.\footnote{Note that the presence of ${\cal B}({\Xi^{*0}_{1530} \to \Xi^{-} \pi^{+}})$ in the denominator of Eq.~\ref{eq:Bres} takes into account the mode $\Xi^{*0}_{1530} \to \Xi^{0} \pi^{0}$, to which this analysis is insensitive.}
The known branching ratios entering into Eqs.~\ref{eq:Bres} and \ref{eq:BR} are ${\cal B}(\Omega^- \to \Lambda K^-)=(67.8\pm0.7$)\%, ${\cal B}(\Xi^- \to\Lambda\pi^-)=(99.887\pm0.035)$\%, ${\cal B}(\Xi^{*0}_{1530} \to \Xi^{-} \pi^{+})=2/3$, and ${\cal B}(K^- \to\pi^-\pi^+\pi^-)=(5.59\pm0.04)$\%~\cite{pdg}. 
The numbers of signal- and normalizing-mode events are obtained from the Gaussian-plus-constant fits of Figs.~\ref{data-minus} and \ref{data-plus}, by subtracting the fitted number of background events from the number of events observed within $\pm3\sigma$ of the mass peak. 
Table~\ref{tab:BR} summarizes the relevant quantities and results.
Since we cannot identify which signal-candidate events under the mass peak are background and which are signal, we performed a study in which randomly chosen candidate events were excluded to see how much the results changed; as expected given the large signal-to-background ratio, the changes were negligibly small.

\begin{table}
\begin{center}
\caption{Direct 3-body and resonance-mode fractions of observed $\Omega^- \rightarrow \Xi^- \pi^+\pi^-$
and ${\overline \Omega}{}^+ \rightarrow {\overline \Xi}{}^+ \pi^+\pi^-$ signals, based on generalized likelihood fits to data.
}
\label{fit-results}
\begin{tabular} {c c c}\hline\hline
Decay & $\Omega^-$ & ${\overline \Omega}{}^+$ \\ 
\hline
$ f_{dir}$ & 0.851$\pm$0.128 & 0.973$\pm$0.245\\
$ f_{res}$ & 0.149$\pm$0.086 & 0.027$\pm$0.144\\
\hline\hline
\end{tabular}
\end{center}
\end{table}

\begin{table}
\begin{center}
\caption{Observed numbers of events, estimated background, acceptances, and measured branching ratios; errors are statistical only.}
\label{tab:BR}
\begin{tabular} {l r @{\,}c@{\,} l r @{\,}c@{\,} l }\hline\hline
Quantity & \multicolumn{3}{c}{$\Omega^-$} &  \multicolumn{3}{c}{${\overline \Omega}{}^+$} \\ \hline
$N^{obs}_{norm}$ & \multicolumn{3}{c}{375}  & \multicolumn{3}{c}{156} \\
$N^{bkg}_{norm}$ & 0.5 & $\pm$ &0.3 & 0.4& $\pm$ &0.2 \\
$N^{obs}_{sig}$  &  \multicolumn{3}{c}{78} &  \multicolumn{3}{c}{24} \\
$N^{bkg}_{sig}$  & 2.3 & $\pm$ & 0.5 & 0.1 & $\pm$ & 0.1 \\
$A_{norm}$ & \qquad \ 2.81& $\times$ & $10^{-4}$ & 2.78& $\times$ &$10^{-4}$ \\ 
$A_{dir}$ & 4.88& $\times$ & $10^{-3}$ & 4.92& $\times$ &$10^{-3}$ \\
$A_{res}$ & 1.20& $\times$ & $10^{-2}$ & \qquad1.20& $\times$ &$10^{-2}$ \\
${\cal B}_{res}$&  \multicolumn{3}{c}{{$(0.40\pm0.24) \!\times\! 10^{-4}$}} &  \multicolumn{3}{c}{{$(0.06\pm0.29) \!\times\! 10^{-4}$}} \\
${\cal B}_{dir}$ &  \multicolumn{3}{c}{{$(3.75\pm0.74) \!\times\! 10^{-4}$}} &  \multicolumn{3}{c}{{$(3.20\pm1.07) \!\times\! 10^{-4}$}}\\
\hline\hline 
\end{tabular}
\end{center}
\end{table}

Systematic uncertainties arise from uncertainties in the known branching ratios of secondary or normalizing decays as reported in \cite{pdg}, variations and uncertainties in the apparatus magnetic fields and in the target and beam locations during the run, and uncertainties in relative geometric acceptance between signal and normalizing modes due to uncertainties in the $\Omega$ momentum spectrum and the $\Omega$, $\Xi$, and $\Lambda$ decay parameters~\cite{pdg}. Although the results of our analysis are dominated by statistical uncertainty, systematic errors were conservatively estimated using the MC, by individually varying the respective parameters by $\pm (\stackrel{>}{_\sim}$$1 \sigma)$, and recalculating the branching ratios with the modified parameter value (details may be found in Ref.~\cite{thesis}).
These error estimates are summarized in Table~\ref{sys}, where (to be conservative) we have symmetrized some asymmetric errors using the worse case.
\begin{table}
\begin{center}
\caption{Systematic errors (in \%) by source and data sample.}
\label{sys}
\begin{tabular} { l r r r r }
\hline \hline
 & \multicolumn{2}{c}{$(\delta {\cal B}/{\cal B})_{dir}$} & \multicolumn{2}{c}{$(\delta {\cal B}/{\cal B})_{res}$}  \\
Source of Error & $\Omega^-$ & ${\overline \Omega}{}^+$ &  $\Omega^-$ & ${\overline \Omega}{}^+$ \\
\hline
Branching-ratio and \\
~~lifetime errors & \raisebox{1.5ex}[0pt]{4.20} & \raisebox{1.5ex}[0pt]{4.20} & \raisebox{1.5ex}[0pt]{3.54} & \raisebox{1.5ex}[0pt]{3.54} \\
MC momentum spectrum & 3.78 & 3.78 & 6.79 & 6.79  \\
Beam targeting & 3.38 & 3.38 & 3.54 & 3.54 \\
Magnetic fields & 1.69 & 1.69 & 1.84 & 1.84 \\
Decay parameters & 1.94 & 1.78 & 2.08 & 1.93 \\
MC statistics & 1.30 & 1.30 & 1.30 & 1.30  \\
\hline
Total (\%) &  7.19 & 7.15 & 8.97 & 8.94 \\
\hline \hline
\end{tabular}
\end{center}
\end{table}
Because this analysis is normalized using a decay mode with five charged particles obtained using the same trigger and offline data set as the signal modes, systematic errors due to such effects as trigger efficiencies tend to cancel and are therefore neglected.

Contrary to theoretical expectation~\cite{fg,atv}, we see little evidence for the resonant decay. To derive an upper limit on the resonant contribution including both statistical and systematic uncertainties, we performed a Monte Carlo simulation of a large sample of hypothetical experiments that took into account the probability distribution from the unbinned maximum-likelihood fit as well as the uncertainty in the branching-ratio normalizing factor (treated as Gaussian-distributed). We find\begin{trivlist}
\item[~~~] ${\cal B}(\Omega^- \to \Xi^{*0}_{1530} \pi^-)<7.4\times10^{-5}\,,$
\item[~~~] ${\cal B}({\overline\Omega}{}^+ \to {\overline  \Xi}{}^{*0}_{1530} \pi^+)<6.2\times10^{-5}\,,$ and 
\item[~~~] ${\cal B}(\Omega^{\mp}\to \Xi^{*0}_{1530}({\overline  \Xi}{}^{*0}_{1530}) \pi^\mp)<7.0\times10^{-5}$
\end{trivlist}  at 90\% confidence level (C.L.). 

The total branching ratio for $\Omega^-$ decaying to $\Xi^-\pi^+\pi^-$ was calculated as a sum of ${\cal B}_{dir}$ and ${\cal B}_{res}\times {\cal B}(\Xi^{*0}_{1530} \to \Xi^{-} \pi^{+})$. Monte Carlo simulation was employed (as described above) to combine statistical and systematic errors taking correlations into account. Median values together with the $\pm$34.1-percentile deviations for the resulting asymmetric distributions of branching ratios are ${\cal B}(\Omega^-\to\Xi^-\pi^+\pi^-) = 4.04^{+0.83}_{-0.71}  \times 10^{-4}$ and ${\cal B}({\overline \Omega}{}^+\to{\overline \Xi}{}^+\pi^+\pi^-) = 3.15^{+1.12}_{-0.89}  \times 10^{-4}$. We see no evidence for {\em CP} violation:  particle and antiparticle branching ratios are consistent with each other.  
Defining the {\em CP} asymmetry 
\begin{equation}A\equiv\frac{{\cal B}(\Omega^-\to\Xi^-\pi^+\pi^-) -{\cal B}({\overline \Omega}{}^+\to{\overline \Xi}{}^+\pi^+\pi^-)}{{\cal B}(\Omega^-\to\Xi^-\pi^+\pi^-) +{\cal B}({\overline \Omega}{}^+\to{\overline \Xi}{}^+\pi^+\pi^-)}\,,
\end{equation}
we find $A=0.12\pm 0.20$.
Combining results for particles and antiparticles, we obtain an average branching ratio ${\cal B}(\Omega^\mp\to\Xi^\mp\pi^+\pi^-) = (3.74 ^{+0.67}_{-0.56}) \times 10^{-4}$. In contrast to the expected $\frac{3}{2}$, this implies ${\cal B}(\Omega^{\mp} \to \Xi^{*0}_{1530}({\overline  \Xi}{}^{*0}_{1530}) \pi^{\mp})/{\cal B}(\Omega^\mp\to\Xi^\mp\pi^+\pi^-)<0.20$ at 90\% C.L. 

In conclusion, using a 25-times-larger sample than that of the previous measurement \cite{wa2}, an improvement in precision has been achieved in determining the branching ratio for $\Omega^- \to \Xi^- \pi^+ \pi^-$. Our observations
of the $\Omega^-$ and $\overline{\Omega}{}^+$ modes agree with {\em CP} conservation and
the final combined branching ratio is $(3.74 ^{+0.67}_{-0.56}) \times 10^{-4}$.
Contrary to expectations, we see no significant contribution of the resonance decay $\Omega^- \to \Xi^{*0}_{1530} \pi^-$ and set a combined upper limit ${\cal B}[\Omega^-({\overline\Omega}{}^+) \to \Xi^{*0}_{1530}({\overline  \Xi}{}^{*0}_{1530}) \pi^\mp]<7.0\times10^{-5}$ at 90\% C.L. 

We wish to thank the Fermilab, LBNL, and university technical staffs for their able 
assistance in commissioning and running this experiment. This work was
supported by the U.S. Dept.\ of Energy and the National Science Council of
Taiwan, R.O.C.


\begin{thebibliography}{99}
\bibitem{antiparticles}All analysis steps were performed identically for positive as for negative data; antiparticles are included when not explicitly stated.
\end{thebibliography}
\end{document}